\documentstyle[aps,multicol,epsf]{revtex}
\begin{document}
\draft
\protect\title{Thermal analysis of hadron multiplicities from 
relativistic quantum molecular dynamics}

\author{J. Sollfrank and U. Heinz}
\address{Institut f\"ur Theoretische Physik,
D-93040 Universit\"at Regensburg, Germany}

\author{H. Sorge} 
\address{Department of Physics, State University of New York 
at Stony Brook, Stony Brook, New York 11794}

\author{N. Xu}
\address{Lawrence Berkeley Laboratory, Berkeley, 
California 94720}

\maketitle

\begin{abstract}
 Some questions arising in the application of the thermal model to
 hadron production in heavy ion collisions are studied.  We do so by
 applying the thermal model of hadron production to particle yields
 calculated by the microscopic transport model RQMD(version 2.3).  We study
 the bias of incomplete information about the final hadronic state on
 the extraction of thermal parameters. It is found that the subset of
 particles measured typically in the experiments looks more thermal
 than the complete set of stable particles. The hadrons which show the
 largest deviations from thermal behavior in RQMD(version 2.3) are the
 multistrange baryons and antibaryons. We also looked at the
 influence of rapidity cuts on the extraction of thermal parameters
 and found that they lead to different thermal parameters and larger
 disagreement between the RQMD yields and the thermal model.
\end{abstract}


\begin{multicols}{2}
\section{Introduction}

The hadron production in nuclear collisions at relativistic and 
ultrarelativistic energies is under extensive experimental and
theoretical investigation \cite{Nagamiya98}. One question
of debate is whether hadron production is thermal,~i.e., whether
the hadron multiplicities for a given volume and energy
density are distributed according to maximum entropy, or whether
the particle multiplicity distribution reflects additional
dynamical constraints.

This question is not easy to answer. First the thermal model has
various versions and depending on the additional input the answers
on thermal parameters differ \cite{Sollfrank97}. Some of the
differences in the various models are the number of included 
resonance states \cite{Sollfrank94}, the use of a canonical or 
grand canonical framework \cite{Cleymans96,Becattini98}, 
the inclusion of hard core repulsion \cite{Yen97}, the inclusion of 
strangeness suppression \cite{Rafelski91}, the inclusion of other 
off-equilibrium parametrizations \cite{Letessier98},
or the inclusion of finite size corrections in the density of states 
\cite{BraunMunzinger95}.

The second important source of uncertainty in the judgment of
thermal behavior is the incomplete measurement of the final
hadronic state. Not all particles which are stable with respect to the 
strong interaction can be measured, and all measurements cover only
a finite part of the complete momentum space. Some measurements cover
nearly the full solid angle, e.g., NA49 \cite{Afanasjev96}, 
in symmetric collision systems, but 
some are restricted to only a small kinematic window 
\cite{Boggild95,Alexeev95,Ambrosini97}.

The bias of the restricted information is the goal of our study here.
We use the transport approach RQMD(version 2.3) \cite{Sorge95}
for generating the full momentum space information
on all stable particles. RQMD provides a microscopic  
description of heavy ion collisions which has been quite successful
in predicting most of the observed features over a wide range of 
conditions and is therefore very well suited for our investigations.

\section{Thermal Model}

The parameters of the thermal model may be understood as Lagrangian
multipliers in extremizing the total entropy under certain
constraints. For a given point particle volume $V$ we have as 
parameters the temperature $T$ (the Lagrangian multiplier to 
achieve a certain average energy density) and chemical potentials 
$\mu_{\cal Q}$ (the Lagrangian multipliers to constrain the system 
on average to a conserved total charge ${\cal Q}$). In our case the 
conserved charges are baryon number $B$, strangeness $S$, and electrical 
charge $Q$. Instead of $B$, $Q$, and $S$ one can equivalently conserve
the net number of valence quarks for each flavor, 
$\langle u - \bar{u} \rangle$, 
$\langle d - \bar{d} \rangle$, 
$\langle s - \bar{s} \rangle$, since on the relevant time scale only
strong interactions are important.
The relations between the chemical potentials in both descriptions are
\begin{eqnarray}
\mu_B & = & \mu_u + 2\mu_d, \nonumber \\
\mu_Q & = & \mu_u - \mu_d, \nonumber \\
\mu_S & = & \mu_d - \mu_s \; .
\end{eqnarray}
The corresponding fugacities are defined by
\begin{equation}
\lambda_f = e^{\mu_f/T} \; , \quad \quad f = u,d,s \; .
\end{equation}

We know from pp collisions that strangeness is not in full chemical
equilibrium with the nonstrange hadrons \cite{Becattini97}. 
Therefore we allow for a deviation of strange particles from complete
chemical equilibrium. However, we force the strange 
particles to be in chemical equilibrium
among each other \cite{Rafelski91}. This requirement is equivalent to the 
introduction of the additional constraint that the total
number of strange valence quarks $\langle s + \bar{s} \rangle$ be
conserved on the average. The Lagrangian multiplier for this constraint
--- expressed as a fugacity --- is \cite{Slotta95}
\begin{equation}
\gamma_s = \exp(\mu_{\langle s + \bar{s} \rangle}/T) \; .
\end{equation}
$\gamma_s$ is a measure of the chemical saturation of strangeness
with $\gamma_s < 1$ meaning undersaturation, 
$\gamma_s = 1$ saturation, i.e., chemical equilibrium,
and $\gamma_s > 1$ oversaturation.

In the grand canonical description with the parameters introduced above
the multiplicity $N_i$ of a particle $i$ with mass $m_i$ and spin 
degeneracy $J_i$ is given by
\begin{eqnarray} \label{yields}
  N_i & = & (2J_i+1)\, \frac{VT}{2\pi^2} 
  \sum_{j=1}^{\infty} (\mp 1)^{j+1} \\ 
 && \times \left(\gamma_s^{\langle s + \bar{s}\rangle_i} \;
 \lambda_u^{\langle u - \bar{u}\rangle_i} 
 \lambda_d^{\langle d - \bar{d}\rangle_i} 
 \lambda_s^{\langle s - \bar{s}\rangle_i} \right)^j \nonumber \\
&& \times \, \int_{-\infty}^{\infty} \!\! d m^2 \;
  \delta(m^2 - m_i^2)\theta(m) \;
  \frac{m^2}{j} \; K_2\left(\frac{j m}{T}\right). 
\end{eqnarray}
The $\langle \dots \rangle_i$ denote the corresponding valence quark
contents of particle species $i$. The upper sign is for fermions,
the lower sign for bosons. With the exception of pions we use for all
particles the Boltzmann approximation; i.e., the sum over the index $j$
is truncated at $j=1$.

For broad resonances ($\Gamma > 1$ MeV) we replace the 
$\delta$ function in Eq.~(\ref{yields}) by a Breit-Wigner distribution 
$\delta_{\rm BW} (m^2)$
\begin{eqnarray}\label{width}
&&\delta\left( m^2 - m_i^2\right) 
\theta \left(m\right) 
\longrightarrow \delta_{\rm BW} (m)  \nonumber \\
&&\quad  = \frac{\alpha\,m_i\Gamma_i}{\left(m^2 - m_i^2\right)^2 
+ m_i^2\Gamma^2_i} \;
\theta\left(m - m_{\rm thres}\right)\, 
\theta\left(m_{\rm cut} - m \right), \nonumber \\
&&
\end{eqnarray}
where $\Gamma_i$ is the width of the resonance $i$, $m_{\rm thres}$ 
a threshold for the production of resonances, and $m_{\rm cut}$ an 
upper cutoff. We use $m_{\rm thres} = m_i - 2\Gamma_i$ and 
$m_{\rm cut} = m_i + 2\Gamma_i$. The normalization $\alpha$ is 
determined by $\int dm^2 \; \delta_{\rm BW}(m^2) = 1$.

Equation (\ref{yields}) gives the multiplicity of primary
hadrons and resonances. The index for the particle species $i$
runs over all hadrons of the Review
of Particle Properties \cite{databook} with $m_i \le 1.7$ GeV.
The final hadron yields of stable hadrons include the
decay contributions of the resonances. Unless otherwise stated
we consider all these hadrons as stable which are stable 
with respect to strong decays. An exception is the $\phi$ which
is taken to be stable in order to be in agreement with the
special mode of RQMD used here (see below). The branching ratios are
taken from \cite{databook}, where missing or inaccurate entries
have been completed by the requirement that all branching
ratios for a resonance have to add to 1.

The thermal model used here is in practical
respects very similar to the one in \cite{Becattini98}.
The main difference is the use of the grand canonical
approach for all conserved charges in the study here.
In addition there are 
small differences in the implementation
of particle properties and decays where accurate numbers are
missing in \cite{databook}. 
We use the following values for the assignment of hidden strangeness:
\begin{eqnarray} 
&&\langle s + \bar{s} \rangle_{\eta} = 
\langle s + \bar{s} \rangle_{\eta^\prime}
 = 1, \nonumber \\ 
&&\langle s + \bar{s} \rangle_{\phi} = 
\langle s + \bar{s} \rangle_{f_0(980)} =
\langle s + \bar{s} \rangle_{f_1(1420)}  =  2, \nonumber \\
&&\langle s + \bar{s} \rangle_{f_1(1510)} =
\langle s + \bar{s} \rangle_{f^\prime_2(1525)} =
\langle s + \bar{s} \rangle_{\phi(1680)} = 2.
\end{eqnarray} 

We show in Tables \ref{tab1} and \ref{tab2} a comparison between the 
fits using the fully grand canonical approach and the partly
canonical approach of\cite{Becattini98}. The two additional
chemical potentials in our approach are
determined by the same constraints as used in \cite{Becattini98},
i.e., strangeness neutrality and electric charge conservation: 
\begin{eqnarray}
 && \sum_i S_i N_i = 0 \label{sconstraint}\\ 
 && \sum_i Q_i N_i = \frac{Z}{A} 
    \sum_i B_i N_i \; . \label{qconstraint}
\end{eqnarray}
$Z/A$ is the ratio of charge over mass number of the
colliding nuclei. We have the same number of 
degrees of freedom as in the canonical approach.
We see from Tables \ref{tab1} and \ref{tab2} that the two
approaches are equivalent on the level of a few percent.

\section{Thermal fits to $4\pi$ multiplicities}

We now turn to RQMD(version 2.3) as the source for our particle
multiplicities. The RQMD transport model \cite{Sorge95} is based 
on the excitation and fragmentation of colour strings followed by 
hadronic reinteractions. The version used [RQMD(version 2.3)] contains 
the color rope mechanism \cite{Sorge92}, an important source for the 
production of the strange hadrons which are of special interest in our 
study. It has been shown by comparison with experimental data 
\cite{Sorge97,Hecke98} that RQMD reproduces the main properties 
in hadron production. It is
therefore a reliable model for generating the full momentum space
information of all produced stable hadrons.
In order to get also a $\phi$ multiplicity we treated in this 
investigation the $\phi$ as a stable particle.

We generated about 400 RQMD events each for S+S and Pb+Pb collisions
with a centrality of 5\% of $\sigma_{\rm inel}^{\rm tot}$. 
The error on the input data is very important in the minimization of 
$\chi^2$. Hadron yields from RQMD have only
a statistical error which can be made arbitrarily small if only enough
events are generated. In order to put thermal fits to experimental
data and thermal fits to RQMD yields on the same footing we give the 
multiplicities calculated by RQMD the same relative error as the 
corresponding experimental value. The actual statistical errors from 
400 RQMD events were in nearly all cases smaller than the corresponding 
experimental ones. We show in 
Table \ref{relerror} a summary of relative errors used for RQMD 
multiplicities, either taken directly from experiment or related to 
other entries of experimental relative errors.

In Tables \ref{S5stat} and \ref{Pb5stat} (rightmost column) the 
$4\pi$ yields of the RQMD(version 2.3) simulation  are shown
as well as yields counted in a central rapidity window of varying size.
The given multiplicities do not contain any decay contributions from
weak or electromagnetic decays. The cited RQMD multiplicities differ
in this respect from the experimental ones. In the thermal fits
the corresponding changes have always been incorporated by the use
of the appropriate decay channels.

We performed different thermal fits to the RQMD $4\pi$ multiplicities.
First, we included in the thermal fit only the subset of particles as 
used in the thermal model description of experimental data in
\cite{Becattini98}.  The result is shown in Table
\ref{rqmd1}. Compared to the thermal fit of experimental data in 
\cite{Becattini98} (see also Table \ref{tab2}) we see that the thermal 
fit parameters of the RQMD analysis are close to the experimental
ones only in the case of S+S collisions. A significant discrepancy 
is seen for the Pb+Pb collisions. The extracted temperature for example
differs by about 30 MeV. We interpret this as an indication of larger
differences in the hadron production between RQMD and the experiment.
We do not try to investigate where the main differences between RQMD 
and the data lie, since this is, first of all, out of scope of this 
paper and, second, most of the experimental data are preliminary and 
thus it is too premature to speculate about physical reasons. The 
reader should further note that a direct comparison of 
experimental yields and the here calculated yields
from RQMD is not reasonable since the yields
of $K$, $\Lambda$, and $\bar{\Lambda}$ contain no weak or
electromagnetic contribution in the case of RQMD while such decays
are partly included in the data. In addition, the
experimental data of Pb+Pb collisions contain hadron
yields and ratios which were measured in a limited acceptance
and were extrapolated to $4\pi$. For all these reasons  
a thermal fit to the experimental Pb+Pb data would be premature.

We would also like to draw the attention to the very low
$\chi^2/{\rm DOF}$ compared to the experimental fit. The small 
value of $\chi^2/{\rm DOF}$ is partly a result of the 
used relative errors for RQMD. Since this error was taken from
experiment, it contains the systematic error of the corresponding 
experiment. As a result of ``perfect reconstruction efficiency" in RQMD 
the error is overestimated. Using only the statistical error of RQMD
would lead to a $\chi^2/{\rm DOF} = 2.37/4$ for S+S collisions
and $\chi^2/{\rm DOF} = 28.00/3$ for Pb+Pb collisions. Nevertheless,
we keep the relative experimental error for RQMD yields, since we 
want to have the same relative weights of hadrons in a thermal fit
as it is typical for an analysis of experimental data. We discuss
the thermal character of RQMD yields in more detail below.

We now show in the rightmost column of Tables \ref{S5statsum} and 
\ref{Pb5statsum} the resulting thermal parameters of a thermal fit 
including all stable particles from RQMD. The comparison between the 
RQMD yields and the thermal ones is given in Tables \ref{S5stat} 
and \ref{Pb5stat}, respectively.
The most striking difference to the fit with the reduced number of
hadrons is the increase in the $\chi^2/{\rm DOF}$ by a factor
of about 25, while the thermal parameters are basically the same.
Only for $\gamma_s$ do we observe a small decrease with increasing
number of stable particles. The result is understood by looking
at the largest deviations of the thermal fit from the 
RQMD yields in Tables \ref{S5stat} and \ref{Pb5stat}.
In terms of standard deviations (s.d.) between the
$4\pi$ thermal fit and RQMD we have the largest deviations for 
$\Omega$ (S+S, 29 s.d.; Pb+Pb, 15 s.d.), 
$\bar\Omega$ (S+S, 9 s.d.; Pb+Pb, 8 s.d.), 
$\phi$ (S+S, 6.7 s.d.; Pb+Pb, 4.7 s.d.) and K$^+$ (S+S, 9 s.d.). 
Interestingly, also in the thermal fit of Ref.
\cite{Letessier98} to the Pb+Pb data the $\bar\Omega$ and $\Omega$
spoil the otherwise good fit, although their deviations go into the
opposite direction.

\begin{figure}[t]
\caption[]{Thermal fit of RQMD $4\pi$ yields of S+S collisions. 
The solid squares denote the input yields of the reduced fit of 
Table \ref{rqmd1}. The open circles show the resulting other 
$4\pi$ multiplicities of stable particles not included in the fit. 
\label{figure1}}
\hspace{0cm}\epsfxsize 15cm \epsfbox{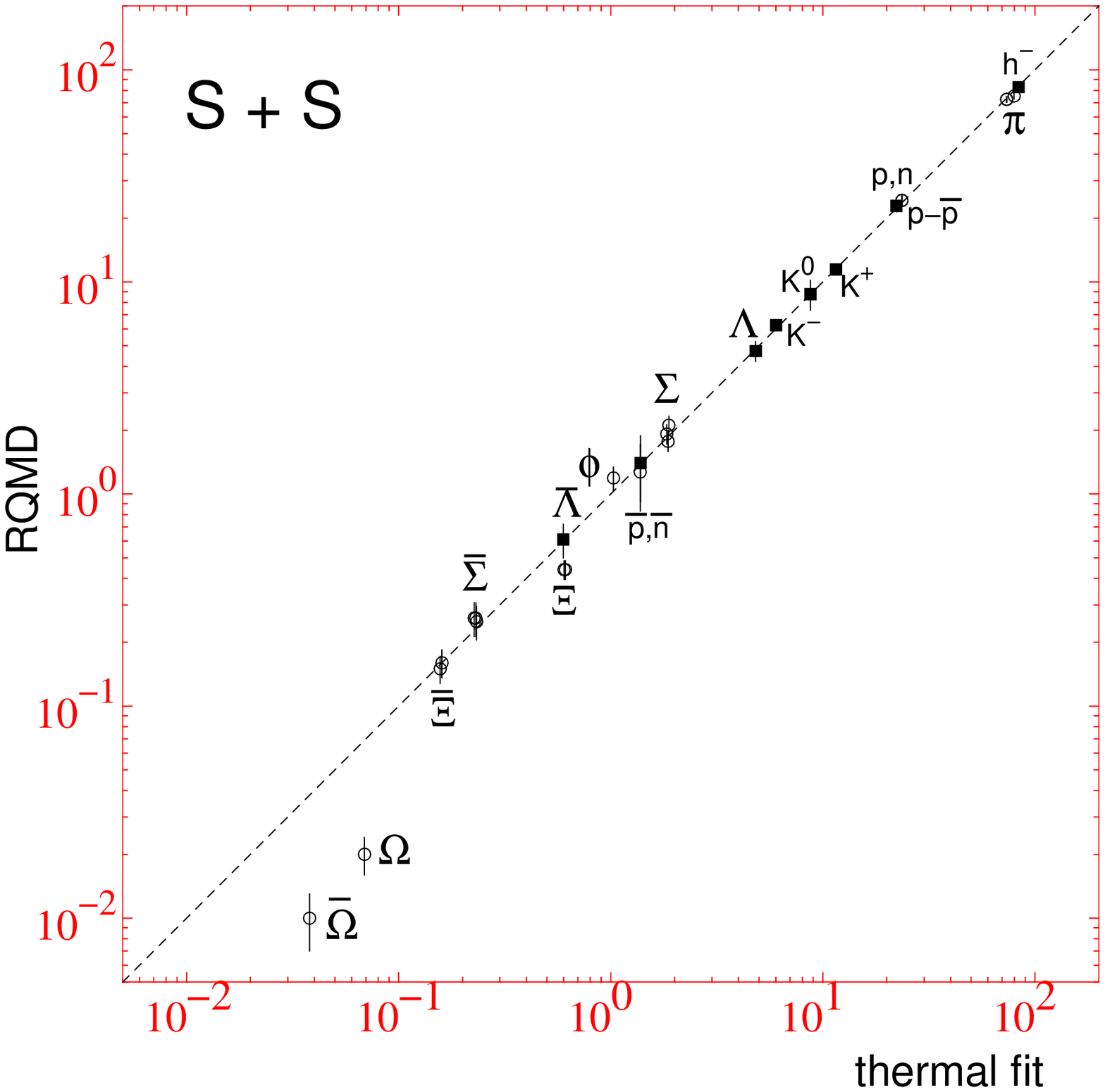}
\end{figure}

\begin{figure}[t]
\caption[]{Thermal fit of RQMD $4\pi$ yields of Pb+Pb collisions. 
The solid squares denote the input yields of the reduced fit of 
Table \ref{rqmd1}. The open circles show the resulting other 
$4\pi$ multiplicities of stable particles not included in the fit. 
\label{figure2}}
\hspace{0cm}\epsfxsize 15cm \epsfbox{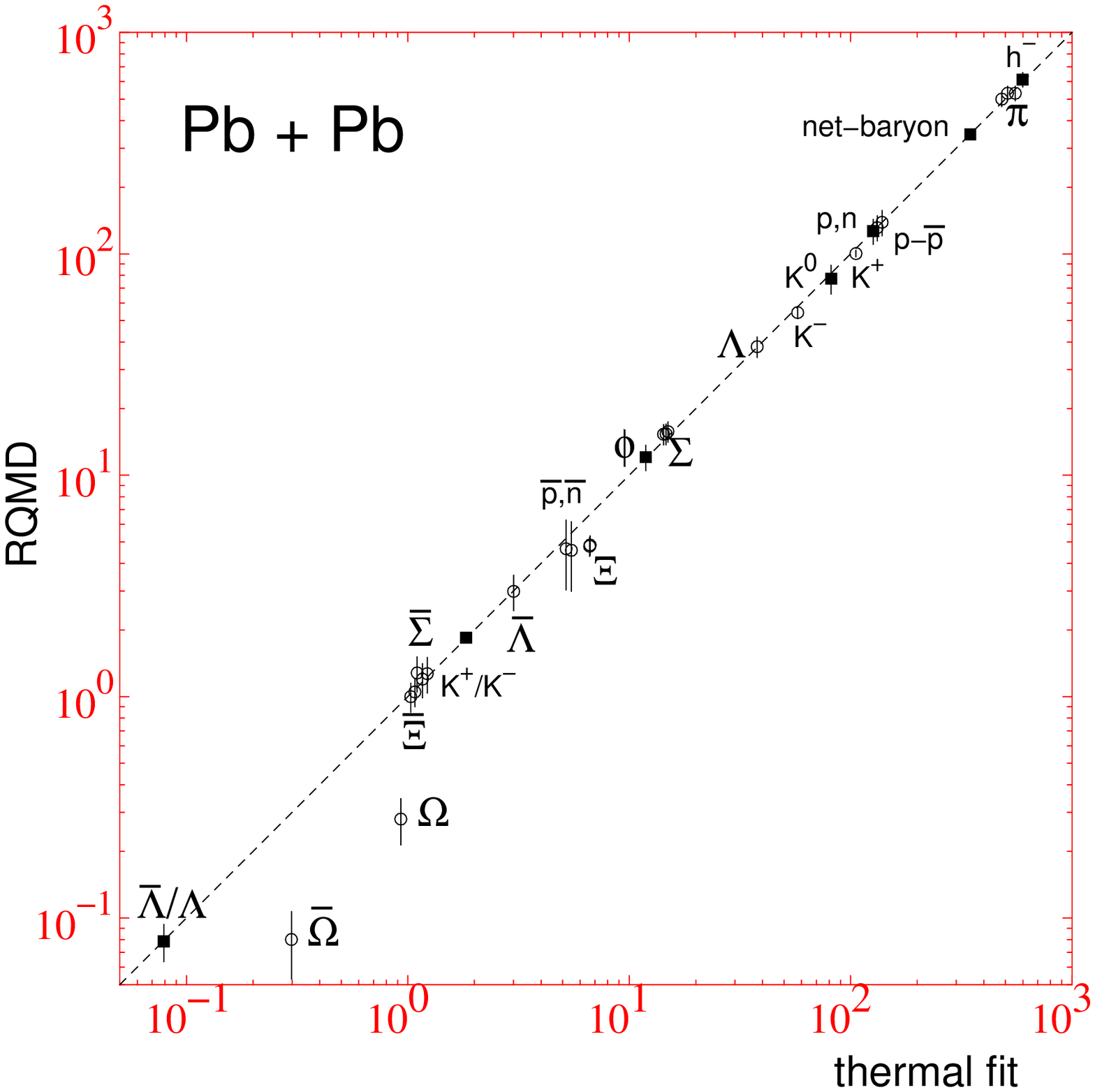}
\end{figure}

The above result shows that the thermal fit gets worse if all 
stable particles are considered, while a reduced number of particle 
species suggest a better thermal behavior. 
The multistrange baryons are mostly responsible for the breakdown
of the otherwise nice thermal behavior. This is illustrated in 
Figs. \ref{figure1} and 
\ref{figure2} where the result of the thermal fit with the reduced
number of input particles of Table \ref{rqmd1}
is displayed. In addition to the fitted input particles 
(solid squares) the resulting other multiplicities of these thermal 
fits are shown as open circles. For both collision systems the 
$\Omega$, $\bar\Omega$, and $\Xi$ show the largest deviations. The
$\chi^2/{\rm DOF}$ of the thermal model with the parameters of 
Table \ref{rqmd1} applied to all stable multiplicities of Tables 
\ref{S5stat} and \ref{Pb5stat}, respectively, 
leads to $\chi^2/{\rm DOF} = 224/21$ for S+S and 
$\chi^2/{\rm DOF} = 191/21$ for Pb+Pb, while excluding 
$\Omega$, $\bar\Omega$, 
$\Xi^0$, $\Xi^-$, $\bar\Xi^0$, and $\bar\Xi^-$ leads to 
$\chi^2/{\rm DOF} = 8.5/15$ for S+S and to 
$\chi^2/{\rm DOF} = 6.9/15$ for Pb+Pb.

We see here the breakdown of the concept of relative strangeness 
equilibrium \cite{Rafelski91} which is violated by various strange 
particle species. The thermal fit has
the tendency to underestimate the meson yields, and the thermal $|S|=1$
and $|S|=2$ (anti)baryons are roughly in agreement with RQMD,
while the $|S|=3$ (anti)baryons are overpredicted by the thermal model
with optimized parameters relative to RQMD. The decrease of $\gamma_s$ 
from the thermal fit with a reduced number of hadrons compared
to the fit with all stable hadrons is a result of the 
above-mentioned undersaturation of the RQMD multistrange hadrons.
This can nicely be seen in Figs. \ref{figure1} and \ref{figure2}.

An explanation for the above findings was offered in 
\cite{Hecke98} where the unexpectedly small inverse slope 
\cite{Andersen98} of the $\Omega$ $p_T$ spectrum was investigated
within the RQMD model. It was found that the $\Omega$ decouples very
early from the rest of the fireball. It is therefore very unlikely
that the $\Omega$ comes into chemical equilibrium with the
rest of the hadrons. The same may also partly be true for the $\Xi$.

A word of caution is necessary here. We see in Figs. \ref{figure1} 
and \ref{figure2} that the $\Omega$ and $\bar\Omega$ have the 
largest deviations from the thermal fit. However, one must note 
that the physics of the $\Omega$($\bar\Omega$) production in RQMD 
is not treated on the same
high quality level as the other particles. Since the $\Omega$ was
of minor physical relevance before --- this has changed with the
high precision measurement of $\Omega$ ratios \cite{Holme97} and 
slopes \cite{Andersen98} --- a precise implementation of the $\Omega$ 
production physics in RQMD was neglected. 
The used strangeness suppression factors and decuplet-octet
suppression factors result already in $e^+ e^-$ collisions in 
$\Omega$ and $\bar\Omega$ yields which are a factor of 2 too low. 
So far no $\Omega$($\bar\Omega$)
production via resonance states has been incorporated. The 
above-mentioned shortcomings reduce the $\Omega$($\bar\Omega$) yields.
Therefore an improvement of the $\Omega$ physics within RQMD will
lead to smaller deviations from the thermal fit than seen now.

We conclude from this section that a relatively small subset 
of hadrons looks more thermal than the complete set of stable 
particles. Thermal behavior should therefore always be tested 
taking as many hadron species into account as possible. 
Nevertheless, we saw only minor changes in the thermal parameters 
between the thermal fit with a reduced number of hadrons
and the thermal fit with all stable hadrons. Therefore, the 
experimentally accessible subset of hadrons is suited to 
determine the thermal parameters temperature, chemical potentials
and the fireball point particle volume. 
$\gamma_s$, however, is sensitive (at least in 
the case of RQMD) to the multistrange (anti)baryons and should 
therefore be extracted by considering the full set of particles.

\section{Thermal fits including rapidity cuts}

Particle yields or ratios are very often only accessible in a
limited kinematic range. While an extrapolation in $p_T$ can be 
performed rather reliably due to the observed nearly exponential 
behavior of the $m_T$ spectra, an extrapolation in rapidity is 
much more difficult 
because different hadron species are known to have quite differently 
shaped rapidity spectra. A thermal fit in case of kinematic cuts
needs model assumptions about the momentum distributions of the 
particles. The simplest assumption one can make is the assumption 
of Bjorken scaling in rapidity \cite{Bjorken83}. However, as a result
of the much too low beam energy, Bjorken scaling has not been seen 
in RQMD or in the experimental data at SPS energies. Taking, 
however, the analyzed rapidity window to be very narrow around 
midrapidity the Bjorken scaling assumption
might be approximately valid. In \cite{Sollfrank98b} it is shown
that within the hydrodynamic model the extraction of thermal
parameters from yields in a limited rapidity range works quite 
reasonably. We like to investigate the systematic error of 
rapidity cuts in a similar way with the help of RQMD events.

For this investigation we first integrate the particle momentum
distribution over 
various rapidity intervals (see, e.g., Table \ref{Pb5stat}). 
Then we assume Bjorken scaling around midrapidity in the sense
that these yields or ratios are fitted by Eq.~(\ref{yields}) 
without correction due to the finite rapidity interval.
Therefore the fitting procedure is the same as for the
full momentum space yields, but the volume has to 
be considered as the sum of comoving subvolumes.

We show in Tables \ref{S5stat} and \ref{S5statsum} the results 
for central S+S collisions and in Tables \ref{Pb5stat} and 
\ref{Pb5statsum} the corresponding results for Pb+Pb collision.
In S+S collisions we see a clear decrease of the temperature
with decreasing width of the rapidity window which is accompanied 
by an increase in $\gamma_s$ and
a decrease in $\lambda_{u,d}$. The decrease of $\lambda_{u,d}$ is
a result of the baryon hole around midrapidity in S+S collisions
\cite{Bachler94}. 
The increase in $\gamma_s$ reflects the fact that RQMD, like the
experiment, produces strange quarks more efficiently in the central
region. On a weaker scale the same tendencies are seen for Pb+Pb 
collisions. 

An interpretation of this result is done in connection with
a similar study of S+S collisions within the hydrodynamic model 
\cite{Sollfrank98b}. In this study 
much smaller deviations between a global thermal fit and particle
yields resulting from integration over different rapidity windows 
were found. In order to understand the difference one has to realize
that in RQMD the physics, especially the physics of strange particle
production, in the central region is different from the fragmentation
regions. Therefore the change here in thermal parameters with the 
analyzed rapidity window resembles mostly a change in physics, 
while in the case of the hydrodynamical model by construction 
the pure kinematic bias was investigated. While a purely kinematic 
influence increases the temperature in the central region 
\cite{Sollfrank98b} the differences in physics decrease the central 
temperature compared to a $4\pi$ analysis. It was further found 
\cite{Sollfrank98b} that for systems with large longitudinal flow 
the artificial change of thermal parameters is small. Therefore 
the change of thermal parameters with the rapidity window in the 
case of RQMD is clearly dominated by the physics component. 

Thus we interpret the large change of thermal parameters with 
different rapidity window size as a result of no global equilibrium 
in these collisions and the deviations from a global equilibrium 
is much larger in S+S collisions compared to Pb+Pb collisions. 
However, there might still be local thermal and chemical
equilibrium in the experiments as well as in the RQMD model.
For a study of local equilibrium in RQMD we refer to \cite{Sorge96}
and in UrQMD to \cite{Bravina98}. We only want here to emphasize that
rapidity cuts can lead to considerable changes in the thermal 
parameters adjusted to a collision system. The change is the
larger the further away the system is from global equilibrium.
In addition we see that the $\chi^2/{\rm DOF}$ is best for $4\pi$ 
integrated input data; as the rapidity interval is narrowed, we see a 
fluctuating but increasing $\chi^2/{\rm DOF}$.

\section{Conclusions}

 We have used the RQMD model to address some questions arising in the
 application of the thermal model to particle production in
 ultrarelativistic heavy ion collisions.  We have performed a thermal
 fit to RQMD multiplicities, where the errors in the RQMD yields have
 been adjusted to be of the same order as typical for
 experiments. First we found that for an input consisting of only a
 small number of hadron species --- typical for existing experiments ---
 the RQMD yields can be nicely reproduced by a thermal model. This
 suggests at a first glance global thermal behavior. However,
 comparing the thermal model with the full set of stable particles
 leads to significant deviations.  The largest discrepancy from
 thermal behavior is found for the multistrange hadrons.  The usual
 small sample of particle species still gives nearly the same thermal
 parameters as the fit with the full set of stable hadrons.
 Experimentally the $4\pi$ data are very scare and extrapolation is always
 required when a thermal fit is performed. Our results imply that one
 must be careful when using the thermal model to analyze the
 experimental data since the constrain is global, {\it not local} 
 \cite{nxu98}.

 It is difficult to assess whether the complete set of $4\pi$
 multiplicities including the multistrange (anti)particles from RQMD
 shows still ``good" thermal behavior. A judgment on the basis of the
 absolute value of $\chi^2/{\rm DOF}$ is not reasonable because
 $\chi^2/{\rm DOF}$ depends strongly on the assumed error of the
 multiplicities and, given certain existing deviations of the RQMD
 output from thermal behavior, becomes arbitrarily large as the
 statistical error on the yields becomes smaller and smaller by
 generating more and more events.  A more useful criterion for judging
 the quality of the thermal fits may be given by the average deviation
 of the best thermal fit (with the relative errors from experiment)
 from the RQMD mean multiplicity.  It is of the order of 20\% for the
 fit in Tables \ref{S5stat} and \ref{Pb5stat}, of order 8\% for a fit
 excluding the $\Omega$ and $\bar\Omega$, and of order 4\% for a fit
 excluding $\Omega$, all $\Xi$ states and the corresponding
 antiparticles.  Given that the absolute yields of the different
 hadrons species vary over three orders of magnitude these numbers
 suggest that RQMD yields are roughly in agreement with a thermal
 model interpretation. However, it must be shown --- similar to the
 studies for the kinetic freeze-out in RQMD \cite{Sorge96} 
 or as done for the UrQMD model in \cite{Bravina98} --- that the
 extracted chemical freeze-out parameters are meaningful in the sense
 that the microscopic state of RQMD at chemical freeze-out is in
 agreement with chemical and kinetic equilibrium and that the
 resulting microscopic thermal parameters are the same as for the
 analysis presented here.

In a very recent publication the result of a thermal fit
to UrQMD midrapidity ratios was presented \cite{Bass98}. In the
case of Pb + Pb $158A$ GeV collisions an agreement was found 
between the UrQMD multiplicities and the thermal model 
on the same quality level as presented here but 
with slightly different parameters
for temperature 140 MeV\cite{Bass98} vs 154 MeV here 
(see Table \ref{Pb5statsum} left column) and 
$\mu_B/T = 1.50$\cite{Bass98} vs 1.57 here.
The detailed microscopic study in \cite{Bass98} revealed that two
fundamental assumptions of the ``thermal model," i.e., universal
freeze-out and global equilibrium, are violated; from this the authors
\cite{Bass98} concluded that thermal fits to particle ratios are
meaningless. We, on the other hand, would like to reserve judgement 
on the meaningfulness of thermal fits to particle multiplicities
until we understand why a thermal fit to particle yields 
works much better than the large violation of common freeze-out
\cite{Hecke98,Bass98} and global equilibrium \cite{Bass98,Bravina98} 
would suggest.

 We also investigated the influence of rapidity cuts on the extracted
 thermal parameters. There is a change of thermal parameters with
 decreasing size of the analyzed rapidity window.  This change is
 dominantly due to nonglobal equilibrium effects in these
 collisions. Thus one has to keep in mind that there is a nonnegligible 
 bias on the extracted thermal parameters due to cuts in rapidity.

\noindent
{\bf Acknowledgments}\\
We would like to thank S.A. Bass for fruitful discussions.
J.S., U.H., and N.X. acknowledge the hospitality of the INT (Seattle)
where part of the work was done. The work was supported by BMBF, DFG,
and GSI.

\newcommand{\IJMPA}[3]{{ Int.~J.~Mod.~Phys.} {\bf A#1}, #3 (#2)}
\newcommand{\JPG}[3]{{ J.~Phys. G} {\bf {#1}}, #3 (#2)}
\newcommand{\AP}[3]{{ Ann.~Phys. (NY)} {\bf {#1}}, #3 (#2)}
\newcommand{\NPA}[3]{{ Nucl.~Phys.} {\bf A{#1}}, #3 (#2)}
\newcommand{\NPB}[3]{{ Nucl.~Phys.} {\bf B{#1}}, #3 (#2)}
\newcommand{\PLB}[3]{{ Phys.~Lett.~B} {\bf {#1}}, #3 (#2)}
\newcommand{\PRv}[3]{{ Phys.~Rev.} {\bf {#1}}, #3 (#2)}
\newcommand{\PRC}[3]{{ Phys.~Rev. C} {\bf {#1}}, #3 (#2)}
\newcommand{\PRD}[3]{{ Phys.~Rev. D} {\bf {#1}}, #3 (#2)}
\newcommand{\PRL}[3]{{ Phys.~Rev.~Lett.} {\bf {#1}}, #3 (#2)}
\newcommand{\PR}[3]{{ Phys.~Rep.} {\bf {#1}}, #3 (#2)}
\newcommand{\ZPC}[3]{{ Z.~Phys. C} {\bf {#1}}, #3 (#2)}
\newcommand{\ZPA}[3]{{ Z.~Phys. A} {\bf {#1}}, #3 (#2)}
\newcommand{\JCP}[3]{{ J.~Comp.~Phys.} {\bf {#1}}, #3 (#2)}
\newcommand{\HIP}[3]{{ Heavy Ion Physics} {\bf {#1}}, #3 (#2)}
\newcommand{\EPJC}[3]{{ Eur.~Phys. J.~C} {\bf {#1}}, #3 (#2)}
\newcommand{\etal}{{\it et al.}}

\end{multicols}

\clearpage
\tighten

\begin{table}
 \caption[]{Comparison between fitted and measured hadron abundances 
  and ratios. Both thermal model calculations include the 
  weak/electromagnetic decay of $\Sigma^0$, $\Xi^0$, $\Xi^-$, 
  $\Omega$ and the corresponding antiparticles.
  \label{tab1}}

   \begin{tabular}{lcccc} 
  Hadron                  & Measured           
& Ref. & Fitted in \cite{Becattini98}& Grand canonical fit \\ \hline
\multicolumn{5}{c}{S+S collisions}\\ \hline
   h$^-$  ($^{a}$)         & 98$\pm$3      &  \cite{Alber98}    
&  92.63     &  94.21     \\ 
   K$^+$                   & 12.5$\pm$0.4  &  \cite{Bachler93}    
&  12.68     &  12.65     \\ 
   K$^-$                   & 6.9$\pm$0.4   &  \cite{Bachler93}    
&  7.611     &  7.27     \\ 
   K$^0_s$                 & 10.5$\pm$1.7  &  \cite{Alber94}    
&  9.939     &  9.72     \\ 
   $\Lambda$  ($^{b}$)     & 9.4$\pm$1.0   &  \cite{Alber94}    
&  7.692     &  8.30     \\ 
$\bar\Lambda$ ($^{b}$)     & 2.2$\pm$0.4   &  \cite{Alber94}    
&  1.474     &  1.633     \\ 
 p-$\bar{\rm p}$ ($^{c}$)  & 21.2$\pm$1.3  &  \cite{Alber98}    
&  21.49     &  22.47     \\ 
 $\bar{\rm p}$  ($^{d}$)   & 1.15$\pm$0.4  &  \cite{Alber96}    
&  2.092     &  2.044     \\
    \hline
\multicolumn{5}{c}{Pb+Pb collisions}\\ \hline
  Net baryon               & 372$\pm$10    &  \cite{Huang97} 
& 375.7    & 374.3    \\ 
   h$^-$         ($^{a}$)  & 680$\pm$50    &  \cite{Afanasjev96} 
& 650.2    & 661.0     \\ 
   K$^0_s$                 & 68$\pm$10     &  \cite{Afanasjev96} 
& 58.27    & 58.22    \\ 
   $\phi$                  & 5.4$\pm$0.7   &  \cite{Friese97} 
& 5.759    & 5.660    \\ 
 p-$\bar{\rm p}$ ($^{c}$)  & 155$\pm$20    &  \cite{Afanasjev96} 
& 155.3    & 152.8    \\ 
   K$^+$/K$^-$             & 1.8$\pm$0.1   &  \cite{Bormann97} 
& 1.652    & 1.699    \\ 
$\bar\Lambda/\Lambda$($^{b}$)& 0.2$\pm$0.04&  \cite{Bormann97} 
& 0.188    & 0.195     \\
   \hline
\multicolumn{5}{l}{$a$ - Defined as $\pi^- + {\rm K}^- + 
{\rm{\bar p}}$.}\\
\multicolumn{5}{l}{$b$ - Includes feeding from $\Xi$.}\\
\multicolumn{5}{l}{$c$ - Measured with the `+' -- `--' method, 
in this case limited}\\
\multicolumn{5}{l}{      rapidity acceptance (0.2-5.8) to 
exclude spectators.}\\
\multicolumn{5}{l}{$d$ - Measured in a restricted rapidity 
interval and extrapolated.}\\
\multicolumn{5}{l}{      by assuming that ${\rm{\bar p}}$ 
has the same rapidity distribution
                           as the $\bar\Lambda$.}\\
\end{tabular}
\end{table}

\vspace{3cm}
\begin{table}[htb] 
 \caption[]{Hadron gas model fitted parameters. 
  The fit itself is shown in Table \ref{tab1}. The entries denoted
  by an asterisk are derived quantities while the others are fit 
  parameters used by MINUIT. \label{tab2}}
 
\begin{tabular}{ccccc }
             &  \multicolumn{2}{c}{S+S} & \multicolumn{2}{c}{Pb+Pb} \\
\cline{2-3}\cline{4-5}
  parameter  &  ref.\cite{Becattini98} &  grand canonical 
             &  ref.\cite{Becattini98} &  grand canonical \\
\hline
      $T$ (MeV)                 & 182.4$\pm$9.2   & 187.7$\pm$1.4  
                                & 192.6$\pm$8.1   & 188.8$\pm$1.1\\
  $\!VT^3e^{-0.7 {\rm GeV}/T}\!$& 3.51$\pm$0.15   &  $3.42^{*}$  
                                & 24.3$\pm$1.6    &  $24.4^{*}$ \\
  $V$ (fm$^3$)                  &  $206^{*}$      & 166$\pm$11 
                                &  $992^{*}$      & 1134$\pm$76 \\
      $\gamma_s$                & 0.732$\pm$0.038 & 0.731$\pm$0.033 
                                & 0.616$\pm$0.043 & 0.620$\pm$0.047 \\
  $\mu_{B}/T$                   & 1.248$\pm$0.074 & $1.251^{*}$ 
                                & 1.207$\pm$0.071 & $1.258^{*}$ \\
  $\lambda_u$                   &  --             & 1.516$\pm$ 0.021
                                &  --             & 1.469$\pm$ 0.018 \\
  $\lambda_d$                   &  --             & 1.518$\pm$ 0.021
                                &  --             & 1.547$\pm$ 0.022\\
  $\lambda_s$                   &  --             & 1.069$\pm$ 0.005
                                &  --             & 1.071$\pm$ 0.005\\ 
\hline
      $\chi^2/$DOF              & 17.1/4          &  12.0/4         
                                & 3.99/3          &  2.33/3 \\ 
\end{tabular}
\end{table}

\clearpage
\begin{table}[htb] 
\caption{List of relative errors of $4\pi$ multiplicities and ratios
used for RQMD multiplicities. \label{relerror}}

\begin{tabular}{cclcl}
& \multicolumn{2}{c}{S+S} & \multicolumn{2}{c}{Pb+Pb}\\ 
\cline{2-3}\cline{4-5}
particle & rel. error in \% & source & rel. error in \% & source \\ 
\hline
h$^-$ & 3.1 & \cite{Alber98} & 7.4 & \cite{Afanasjev96} \\
p-$\bar{\rm p}$ & 6.1 & \cite{Alber98} & 12.9 & \cite{Afanasjev96} \\
net baryon & -- & --  & 2.7 & \cite{Huang97} \\
K$^+$/K$^-$ & --  & -- & 5.6 & \cite{Bormann97} \\
$\bar\Lambda/\Lambda$ & -- & -- & 20.0 & \cite{Bormann97} \\
p & 6.1 & same as p-$\bar{\rm p}$ & 12.9 & same as p-$\bar{\rm p}$ \\
$\bar{\rm p}$ & 34.8 & \cite{Alber96} & 34.8 & same as S+S \\
n & 6.1 & same as p & 12.9 & same as p \\
$\bar{\rm n}$ & 34.8 & same as $\bar{\rm p}$ & 34.8 & same as 
$\bar{\rm p}$ \\
$\pi^+$ & 3.1 & same as h$^-$ & 7.4 &  same as h$^-$ \\
$\pi^-$ & 3.1 & same as h$^-$ & 7.4 &  same as h$^-$ \\
$\pi^0$ & 3.1 & same as h$^-$ & 7.4 &  same as h$^-$ \\
K$^+$ & 3.2 & \cite{Bachler93} & 3.2 & same as S+S \\
K$^-$ & 5.8 & \cite{Bachler93} & 5.8 & same as S+S \\
K$_s^0$ & 16.2 & \cite{Alber94} & 14.7 & \cite{Afanasjev96} \\
$\phi$  & 13.0 & same as Pb+Pb & 13.0 & \cite{Friese97} \\
$\Lambda$ & 10.6 & \cite{Alber94} & 10.6 & same as S+S \\
$\bar{\Lambda}$ & 18.2 & \cite{Alber94} & 18.2 & same as S+S \\
$\Sigma^+$ & 10.6 & same as $\Lambda$ & 10.6 & same as $\Lambda$ \\ 
$\bar{\Sigma}^+$ & 18.2 & same as $\bar\Lambda$ & 18.2 
& same as $\bar\Lambda$ \\ 
$\Sigma^0$ & 10.6 & same as $\Lambda$ & 10.6 & same as $\Lambda$ \\ 
$\bar{\Sigma}^0$ & 18.2 & same as $\bar\Lambda$ & 18.2 
& same as $\bar\Lambda$ \\ 
$\Sigma^-$ & 10.6 & same as $\Lambda$ & 10.6 & same as $\Lambda$ \\ 
$\bar{\Sigma}^-$ & 18.2 & same as $\bar\Lambda$ & 18.2 
& same as $\bar\Lambda$ \\ 
$\Xi^-$ & 10.0 & same as Pb+Pb & 10.0 & \cite{Andersen98}($^a$) \\ 
$\bar{\Xi}^-$ & 15.0 & same as Pb+Pb & 15.0 
& \cite{wa97private}($^b$) \\
$\Xi^0$ & 10.0 & same as $\Xi^-$ & 10.0 & same as $\Xi^-$ \\
$\bar{\Xi}^0$ & 15.0 & same as $\bar\Xi^-$ & 15.0 
& same as $\bar\Xi^-$ \\
$\Omega^-$ & 23.0 & same as Pb+Pb & 23.0 
& \cite{Andersen98}($^a$) \\ 
$\bar{\Omega}^-$ & 34.5 & same as Pb+Pb & 34.5 
& \cite{wa97private}($^b$) \\
\hline
\multicolumn{5}{l}{($^a$) We assume that the relative error
is the same as the error on multiplicity } \\
\multicolumn{5}{l}{ \phantom{($^a$)} per event in the kinematic
window of WA97 for the most central events \cite{Andersen98}.}\\ 
\multicolumn{5}{l}{($^b$) 
The relative error of multiplicities of multistrange antibaryons 
is typically} \\
\multicolumn{5}{l}{ \phantom{($^b$)} 50\% larger
than the relative error of the corresponding baryons \cite{wa97private}.}
\end{tabular}
\end{table}

\clearpage
\begin{table}
\squeezetable
\caption{ RQMD(version 2.3) multiplicities of $200A$ GeV/c S+S collisions. 
Trigger cross section $\sigma_{\rm trig} \leq 5\% \sigma_{\rm geom}$ 
($b \leq 0.85$ fm). \protect\label{S5stat}}

\begin{tabular}{ccccccccccc}
 & \multicolumn{2}{c}{$|y|\leq$0.5}
 & \multicolumn{2}{c}{$|y|\leq$1.0}
 & \multicolumn{2}{c}{$|y|\leq$1.5}
 & \multicolumn{2}{c}{$|y|\leq$2.0}
 & \multicolumn{2}{c}{$4\pi$}\\
\cline{2-3}\cline{4-5}\cline{6-7}\cline{8-9}\cline{10-11}
&RQMD&therm.&RQMD&therm.&RQMD&therm.&RQMD&therm.&RQMD&therm.\\
\hline
$p$           &   3.13$\pm$0.19  & 3.297  &   6.95$\pm$0.42  & 7.304 
              &  11.65$\pm$0.71  & 11.954 &   16.47$\pm$1.00 & 16.624 
              &  24.24$\pm$1.47  & 23.552 \\ 
$\bar{p}$     &   0.50$\pm$0.17  & 0.342  &    0.99$\pm$0.34 & 0.975 
              &   1.25$\pm$0.43  & 1.122  &    1.36$\pm$0.47 & 1.187 
              &   1.40$\pm$0.48  & 1.301 \\ 
$n$           &   3.20$\pm$0.19  & 3.292  &    7.23$\pm$0.44 & 7.296 
              &  11.79$\pm$0.71  & 11.943 &   16.60$\pm$1.01 & 16.611 
              &  24.23$\pm$1.47  & 23.536 \\ 
$\bar{n}$     &   0.49$\pm$0.17  & 0.340  &    0.89$\pm$0.30 & 0.972 
              &   1.12$\pm$0.39  & 1.119  &    1.23$\pm$0.42 & 1.183 
              &   1.27$\pm$0.44  & 1.297 \\ 
$\pi^+$       &  18.42$\pm$0.57  & 18.719 &   35.78$\pm$1.10 & 36.959 
              &  50.36$\pm$1.56  & 51.527 &   60.89$\pm$1.88 & 62.484 
              &  72.54$\pm$2.24  & 74.863 \\ 
$\pi^-$       &  18.80$\pm$0.58  & 18.732 &   36.10$\pm$1.11 & 36.983 
              &  50.56$\pm$1.56  & 51.564 &   60.18$\pm$1.86 & 62.532 
              &  72.69$\pm$2.25  & 74.919 \\ 
$\pi^0$       &  19.18$\pm$0.59  & 20.562 &   37.15$\pm$1.15 & 40.267 
              &  52.31$\pm$1.62  & 56.141 &   63.38$\pm$1.96 & 67.837 
              &  75.31$\pm$2.33  & 80.891 \\ 
$K^+$         &   3.34$\pm$0.10  & 3.136  &    6.40$\pm$0.20 & 5.477 
              &   8.92$\pm$0.28  & 8.018  &   10.51$\pm$0.33 & 9.384 
              &  11.45$\pm$0.36  & 10.354 \\ 
$K^-$         &   2.10$\pm$0.12  & 2.152  &    3.86$\pm$0.22 & 3.702 
              &   5.16$\pm$0.29  & 4.988  &    5.91$\pm$0.34 & 5.469 
              &   6.26$\pm$0.36  & 5.567 \\ 
$K^0_s$       &   2.70$\pm$0.43  & 2.623  &    5.08$\pm$0.82 & 4.563 
              &   6.98$\pm$1.13  & 6.467  &    8.11$\pm$1.31 & 7.386 
              &   8.76$\pm$1.41  & 7.921 \\ 
$\phi$        &   0.42$\pm$0.054 & 0.309  &    0.81$\pm$0.10 & 0.519 
              &   1.06$\pm$0.13  & 0.754  &    1.17$\pm$0.15 & 0.811 
              &   1.19$\pm$0.15  & 0.790 \\ 
$\Lambda$     &   1.05$\pm$0.11  & 0.838  &    2.21$\pm$0.23 & 1.663 
              &   3.32$\pm$0.35  & 2.688  &    4.18$\pm$0.44 & 3.416 
              &   4.72$\pm$0.50  & 4.200 \\ 
$\bar{\Lambda}$&  0.25$\pm$0.045 & 0.131  &    0.44$\pm$0.080& 0.343 
              &   0.56$\pm$0.10  & 0.429  &    0.60$\pm$0.10 & 0.447 
              &   0.61$\pm$0.11  & 0.469 \\ 
$\Sigma^+$    &   0.50$\pm$0.053 & 0.329  &    0.96$\pm$0.10 & 0.650 
              &   1.49$\pm$0.15  & 1.050  &    1.84$\pm$0.19 & 1.333 
              &   2.11$\pm$0.22  & 1.636 \\ 
$\bar{\Sigma}^+$& 0.10$\pm$0.018 & 0.051  &    0.17$\pm$0.030& 0.134 
              &   0.23$\pm$0.041 & 0.168  &    0.25$\pm$0.045& 0.175 
              &   0.25$\pm$0.045 & 0.183 \\ 
$\Sigma^0$    &   0.43$\pm$0.045 & 0.325  &    0.82$\pm$0.086& 0.643 
              &   1.29$\pm$0.13  & 1.039  &    1.59$\pm$0.16 & 1.320 
              &   1.77$\pm$0.18  & 1.620 \\ 
$\bar{\Sigma}^0$& 0.11$\pm$0.020 & 0.050  &    0.19$\pm$0.034& 0.133 
              &   0.24$\pm$0.043 & 0.166  &    0.26$\pm$0.047& 0.173 
              &   0.26$\pm$0.047 & 0.181 \\ 
$\Sigma^-$    &   0.47$\pm$0.049 & 0.319  &    0.94$\pm$0.099& 0.633 
              &   1.39$\pm$0.14  & 1.023  &    1.74$\pm$0.18 & 1.300 
              &   1.92$\pm$0.20  & 1.597 \\ 
$\bar{\Sigma}^-$& 0.11$\pm$0.020 & 0.049  &    0.19$\pm$0.034& 0.130 
              &   0.24$\pm$0.043 & 0.163  &    0.26$\pm$0.047& 0.170 
              &   0.26$\pm$0.047 & 0.179 \\ 
$\Xi^0$       &   0.11$\pm$0.011 & 0.135  &    0.25$\pm$0.025& 0.237 
              &   0.36$\pm$0.036 & 0.378  &    0.43$\pm$0.043& 0.439 
              &   0.44$\pm$0.044 & 0.466 \\ 
$\bar{\Xi}^0$ &   0.07$\pm$0.010 & 0.032  &    0.12$\pm$0.018& 0.075 
              &   0.15$\pm$0.022 & 0.102  &    0.16$\pm$0.024& 0.105 
              &   0.16$\pm$0.024 & 0.105 \\ 
$\Xi^-$       &   0.14$\pm$0.014 & 0.133  &    0.26$\pm$0.026& 0.233 
              &   0.36$\pm$0.036 & 0.372  &    0.43$\pm$0.043& 0.432 
              &   0.44$\pm$0.044 & 0.459 \\ 
$\bar{\Xi}^-$ &   0.06$\pm$0.009 & 0.031  &    0.11$\pm$0.016& 0.074 
              &   0.13$\pm$0.019 & 0.101  &    0.15$\pm$0.022& 0.103 
              &   0.15$\pm$0.022 & 0.103 \\ 
$\Omega^-$    &   0.01$\pm$0.002 & 0.016  &    0.01$\pm$0.002& 0.028 
              &   0.02$\pm$0.004 & 0.044  &    0.02$\pm$0.004& 0.047 
              &   0.02$\pm$0.004 & 0.044 \\ 
$\bar{\Omega}^-$& 0.01$\pm$0.003 & 0.005  &    0.01$\pm$0.003& 0.014 
              &   0.01$\pm$0.003 & 0.020  &    0.01$\pm$0.003& 0.021 
              &   0.01$\pm$0.003 & 0.020 \\
\end{tabular} 
\end{table}

\vspace{3cm}
\begin{table}
\squeezetable
\caption{RQMD(version 2.3) multiplicities of $158A$ GeV/c Pb+Pb 
collisions. Trigger cross section 
$\sigma_{\rm trig} \leq 5\% \; \sigma_{\rm geom}$ 
($b \leq 3$ fm). \label{Pb5stat}}

\begin{tabular}{ccccccccccc}
 & \multicolumn{2}{c}{$|y|\leq$0.5}
 & \multicolumn{2}{c}{$|y|\leq$1.0}
 & \multicolumn{2}{c}{$|y|\leq$1.5}
 & \multicolumn{2}{c}{$|y|\leq$2.0}
 & \multicolumn{2}{c}{$4\pi$}\\
\cline{2-3}\cline{4-5}\cline{6-7}\cline{8-9}\cline{10-11}
&RQMD&therm.&RQMD&therm.&RQMD&therm.&RQMD&therm.&RQMD&therm.\\
\hline
$p$           &  27.06$\pm$3.49  & 32.321 &   55.08$\pm$7.10 & 64.884 
              &  82.51$\pm$10.6  & 96.019 &  105.00$\pm$13.5 & 120.752 
              & 131.40$\pm$17.0  &146.992 \\ 
$\bar{p}$     &   1.68$\pm$0.58  & 2.295  &    3.14$\pm$1.09 & 4.011 
              &   3.92$\pm$1.36  & 5.184  &    4.37$\pm$1.52 & 5.746 
              &   4.58$\pm$1.59  & 6.348 \\ 
$n$           &  28.89$\pm$3.72  & 33.513 &   58.00$\pm$7.48 & 67.476
              &  85.82$\pm$11.1  &100.184 &  109.35$\pm$14.1 & 126.259
              & 138.83$\pm$17.9  &154.054 \\ 
$\bar{n}$     &   1.77$\pm$0.61  & 2.203  &    3.25$\pm$1.13 & 3.839 
              &   4.05$\pm$1.40  & 4.946  &    4.45$\pm$1.54 & 5.471 
              &   4.66$\pm$1.62  & 6.030 \\ 
$\pi^+$       & 144.16$\pm$10.7  &156.050 &  271.73$\pm$20.1 & 291.882
              & 369.94$\pm$27.4  &397.705 &  435.64$\pm$32.2 & 470.565
              & 500.05$\pm$37.0  &534.792 \\ 
$\pi^-$       & 151.11$\pm$11.2  &163.344 &  284.80$\pm$21.1 & 306.556
              & 389.28$\pm$28.8  &419.270 &  460.63$\pm$34.1 & 497.486
              & 531.42$\pm$39.3  &567.035 \\ 
$\pi^0$       & 152.56$\pm$11.3  &176.860 &  287.54$\pm$21.3 & 331.296 
              & 391.99$\pm$29.0  &450.767 &  461.97$\pm$34.2 & 532.048 
              & 531.37$\pm$39.3  &601.890 \\ 
$K^+$         &  31.91$\pm$1.02  & 30.007 &   60.16$\pm$1.92 & 56.971
              &  81.12$\pm$2.59  & 76.911 &   93.69$\pm$2.99 & 88.895
              & 100.33$\pm$3.21  & 95.780 \\ 
$K^-$         &  18.76$\pm$1.08  & 19.155 &   34.72$\pm$2.01 & 35.180 
              &  45.55$\pm$2.64  & 45.836 &   51.51$\pm$2.98 & 51.683 
              &  54.28$\pm$3.14  & 54.358 \\ 
$K^0_s$       &  25.59$\pm$3.76  & 24.488 &   47.92$\pm$7.04 & 45.934 
              &  63.73$\pm$9.36  & 61.239 &   72.75$\pm$10.7 & 70.173 
              &  77.30$\pm$11.4  & 75.0 \\ 
$\phi$        &   4.55$\pm$0.59  & 3.179  &    8.19$\pm$1.06  & 5.999 
              &  10.74$\pm$1.39  & 7.853  &   11.82$\pm$1.53  & 8.654 
              &  12.08$\pm$1.57  & 8.681 \\ 
$\Lambda$     &  10.79$\pm$1.14  & 8.936  &   20.63$\pm$2.18  & 17.769 
              &  29.34$\pm$3.11  & 25.326 &   35.38$\pm$3.75  & 30.367
              &  38.09$\pm$4.03  & 34.206 \\ 
$\bar{\Lambda}$&  1.24$\pm$0.22  & 1.019  &    2.06$\pm$0.37  & 1.829 
              &   2.62$\pm$0.47  & 2.367  &    2.90$\pm$0.52  & 2.569
              &   2.99$\pm$0.54  & 2.696 \\ 
$\Sigma^+$    &   4.30$\pm$0.45  & 3.412  &    8.49$\pm$0.89  & 6.770 
              &  12.20$\pm$1.29  & 9.623  &   14.48$\pm$1.53  & 11.518
              &  15.32$\pm$1.62  & 12.946 \\ 
$\bar{\Sigma}^+$& 0.55$\pm$0.10  & 0.411  &    0.88$\pm$0.16  & 0.740 
              &   1.12$\pm$0.20  & 0.960  &    1.25$\pm$0.22  & 1.043 
              &   1.27$\pm$0.23  & 1.096 \\ 
$\Sigma^0$    &   4.21$\pm$0.44  & 3.465  &    8.50$\pm$0.90  & 6.890 
              &  12.07$\pm$1.27  & 9.817  &   14.35$\pm$1.52  & 11.769 
              &  15.38$\pm$1.63  & 13.250 \\ 
$\bar{\Sigma}^0$& 0.47$\pm$0.085 & 0.395  &    0.83$\pm$0.15  & 0.709 
              &   1.07$\pm$0.19  & 0.918  &    1.17$\pm$0.21  & 0.996 
              &   1.20$\pm$0.21  & 1.045 \\ 
$\Sigma^-$    &   4.45$\pm$0.47  & 3.496  &    8.81$\pm$0.93  & 6.967 
              &  12.35$\pm$1.30  & 9.952  &   14.77$\pm$1.56  & 11.949 
              &  15.76$\pm$1.67  & 13.478 \\ 
$\bar{\Sigma}^-$& 0.49$\pm$0.089 &  0.377 &    0.86$\pm$0.15  & 0.676 
              &   1.14$\pm$0.20  & 0.873  &    1.27$\pm$0.23  & 0.945 
              &   1.28$\pm$0.23  & 0.991 \\ 
$\Xi^0$       &   1.57$\pm$0.15  & 1.532  &    3.01$\pm$0.30  & 3.008 
              &   4.13$\pm$0.41  & 4.114  &    4.66$\pm$0.46  & 4.696 
              &   4.78$\pm$0.47  & 4.880 \\ 
$\bar{\Xi}^0$ &   0.42$\pm$0.063 & 0.295  &    0.72$\pm$0.10  & 0.545 
              &   0.95$\pm$0.14  & 0.707  &    1.04$\pm$0.15  & 0.752 
              &   1.05$\pm$0.15  & 0.751 \\ 
$\Xi^-$       &   1.53$\pm$0.15  & 1.528  &    3.04$\pm$0.30  & 3.006 
              &   4.16$\pm$0.41  & 4.118  &    4.71$\pm$0.47  & 4.706 
              &   4.84$\pm$0.48  & 4.896 \\ 
$\bar{\Xi}^-$ &   0.41$\pm$0.061 & 0.284  &    0.69$\pm$0.10  & 0.524 
              &   0.92$\pm$0.13  & 0.679  &    1.00$\pm$0.15  & 0.722 
              &   1.00$\pm$0.15  & 0.720 \\ 
$\Omega^-$    &   0.10$\pm$0.023 & 0.203  &    0.20$\pm$0.046 & 0.396 
              &   0.26$\pm$0.059 & 0.525  &    0.28$\pm$0.066 & 0.571 
              &   0.28$\pm$0.066 & 0.552 \\ 
$\bar{\Omega}^-$& 0.03$\pm$0.010 & 0.064  &    0.06$\pm$0.020 & 0.122 
              &   0.08$\pm$0.027 & 0.160  &    0.08$\pm$0.027 & 0.167 
              &   0.08$\pm$0.027 & 0.159 \\ 
\end{tabular} 
\end{table}

\clearpage
\begin{table}
\caption{\label{rqmd1}
Thermal parameters of fitting RQMD yields and ratios
in S+S (left) and Pb+Pb (right) collisions. 
The same particles and ratios were taken like in
the fit of \protect\cite{Becattini98}.
The input data set was extracted from the entries
in Tables \ref{S5stat} and \ref{Pb5stat}, respectively.
The relative error is as in Table \ref{tab1}.}

\begin{tabular}{cccccc}
&\multicolumn{2}{c}{S+S} & & \multicolumn{2}{c}{Pb+Pb} \\ 
\cline{2-3}\cline{5-6}
particle&RQMD &thermal& particle & RQMD &thermal \\ 
\hline
$h^-$ &82.98$\pm$2.57& 83.77  &
$h^-$ &612.5$\pm$45.32& 598.87 \\
$K_s^0$ &8.76$\pm$1.42& 8.74 & 
$K_s^0$ &77.31$\pm$11.36& 81.71 \\
$p-\bar{p}$ &22.84$\pm$1.39& 22.22 &
$p-\bar{p}$ &126.82$\pm$16.36& 126.50 \\ 
$K^+$ &11.45$\pm$0.37& 11.53 &
$\phi$ &12.0$\pm$1.57& 11.88 \\ 
$K^-$ &6.26$\pm$0.36& 6.02 & 
$K^+/K^-$ &1.847$\pm$0.104& 1.832 \\ 
$\Lambda$ &4.72$\pm$0.50& 4.82 & 
$\bar{\Lambda}/\Lambda$ &0.0785$\pm$0.016& 0.080 \\ 
$\bar{\Lambda}$ &0.61$\pm$0.11& 0.597 &
net~baryon &346.6$\pm$9.36& 347.19 \\
$\bar{p}$ &1.4$\pm$0.49& 1.38 & & &   \\ \hline 
$\chi^2/$DOF & & 0.83/4 & & & 0.29/3 \\ \hline
$T$(MeV) & & 179.4$\pm$7.5 & & & 156.7$\pm$8.6 \\ 
$V$(fm$^3$) & & 195$\pm$65 & & & 3008$\pm$1780 \\ 
$\gamma_s$ & & 0.769$\pm$0.045 & & & 1.129$\pm$0.17 \\ 
$\lambda_u$ & &1.609$\pm$0.026 && & 1.677$\pm$0.044 \\ 
$\lambda_d$ & & 1.610$\pm$0.026 & & & 1.761$\pm$0.065 \\ 
$\lambda_s$ & & 1.102$\pm$0.032 & & & 1.208$\pm$0.059 \\ 
\end{tabular} 
\end{table}

\vspace{2cm}
\begin{table}[htb]
\caption{\label{S5statsum}
Thermal parameters of fitting RQMD multiplicities in various rapidity
intervals for S+S collisions  (trigger cross
section $\sigma_{\rm trig} \leq 5\% \sigma_{\rm geom}$). 
The fit itself is shown in Table \ref{S5stat}. }

\begin{tabular}{cccccc} 
&{$|y|\leq$0.5}&{$|y|\leq$1.0}&{$|y|\leq$1.5}&{$|y|\leq$2.0}&{$4\pi$}\\ 
\hline
$T$(MeV) & {153.3$\pm$1.3} & {166.6$\pm$3.9} & {167.5$\pm$2.7} 
& {168.5$\pm$2.7} & {172.9$\pm$3.0} \\ 
$V$(fm$^3$) & {147$\pm$8} & {173$\pm$29} & {227$\pm$26} 
& {261$\pm$30} & {260$\pm$32} \\ 
$\gamma_s$ & {0.905$\pm$0.027} & {0.768$\pm$0.022} & {0.793$\pm$0.019}
& {0.748$\pm$0.018} & {0.670$\pm$0.016} \\ 
$\lambda_u$ & {1.460$\pm$0.017} & {1.400$\pm$0.016} & {1.485$\pm$0.015}
& {1.555$\pm$0.017} & {1.623$\pm$0.019} \\
$\lambda_d$ & {1.461$\pm$0.017} & {1.401$\pm$0.016} & {1.487$\pm$0.015}
& {1.556$\pm$0.017} & {1.625$\pm$0.019} \\
$\lambda_s$ & {1.188$\pm$0.007} & {1.124$\pm$0.010} & {1.136$\pm$0.010}
& {1.145$\pm$0.011} & {1.138$\pm$0.013} \\ \hline
$\chi^2$/DOF & {115.32/21} & {152.20/21} & {98.51/21}
& {111.91/21} & {94.72/21} \\
\end{tabular} 
\end{table}

\vspace{2cm}
\begin{table}[htb]
\caption{\label{Pb5statsum}
Thermal parameters of fitting RQMD multiplicities in various rapidity
intervals for Pb+Pb collisions  (trigger cross
section $\sigma_{\rm trig} \leq 5\% \sigma_{\rm geom}$). 
The fit itself is shown in Table \ref{Pb5stat}. }

\begin{tabular}{cccccc} 
&{$|y|\leq$0.5}&{$|y|\leq$1.0}&{$|y|\leq$1.5}&{$|y|\leq$2.0}&{$4\pi$}\\ 
\hline
$T$(MeV) & {152.8$\pm$4.8} & {153.3$\pm$3.3} & {154.4$\pm$2.9}
& {154.4$\pm$2.7} & {155.4$\pm$2.6} \\ 
$V$(fm$^3$) & {1223$\pm$277} & {2215$\pm$347} & {2868$\pm$390}
& {3386$\pm$427} & {3706$\pm$460} \\ 
$\gamma_s$ & {1.019$\pm$0.046} & {1.024$\pm$0.037} & {1.000$\pm$0.033}
& {0.966$\pm$0.031} & {0.900$\pm$0.029} \\ 
$\lambda_u$ & {1.538$\pm$0.032} & {1.573$\pm$0.024} & {1.608$\pm$0.022}
& {1.641$\pm$0.022} & {1.667$\pm$0.022} \\ 
$\lambda_d$ & {1.595$\pm$0.041} & {1.635$\pm$0.031} & {1.676$\pm$0.028}
& {1.714$\pm$0.028} & {1.745$\pm$0.028} \\ 
$\lambda_s$ & {1.211$\pm$0.017} & {1.216$\pm$0.014} & {1.219$\pm$0.014}
& {1.227$\pm$0.014} & {1.230$\pm$0.014} \\ \hline
$\chi^2/$DOF & {79.62/21} & {66.17/21} & {67.38/21}
& {67.61/21} & {55.61/21} \\ 
\end{tabular} 
\end{table}

\clearpage

\clearpage

\end{document}